\documentclass[floatfix,nofootinbib,twocolumn,showpacs,superscriptaddress, reprintnumbers,amssymb,letterpaper,amsmath,pra,amsfonts]{revtex4-2}

\usepackage{graphicx}
\usepackage{dcolumn}
\usepackage{bm}
\usepackage{xcolor}
\usepackage[colorlinks=true,allcolors=blue]{hyperref}
\usepackage{amsmath}
\usepackage{lineno}
\usepackage{comment}
\usepackage{tikz}
\usepackage{url}
\usepackage{cleveref}
\usetikzlibrary{shapes.geometric, arrows}

\begin{document}
\title{Calculation of RF-induced Temporal Jitter in Ultrafast Electron Diffraction}

\author{T. Xu}
\email{txu@slac.stanford.edu}
\author{F. Ji}
\author{S. P. Weathersby}
\author{R. J. England}
\affiliation{SLAC National Accelerator Laboratory, Menlo Park, CA 94025, USA}

\date{\today}

\begin{abstract}
A significant contribution to the temporal resolution of an ultrafast electron diffraction (UED) instrument is arrival time jitter caused by amplitude and phase variation of radio frequency (RF) cavities. In this paper, we present a semi-analytical approach for calculating RF-induced temporal jitter from klystron and RF cavity parameters. Our approach allows fast estimation of temporal jitter for MeV-UED beamlines and can serve as a virtual timing tool when shot-to-shot measurements of RF amplitude and phase jitters are available. A simulation study for the SLAC MeV-UED instrument is presented and the temporal resolution for several beamline configurations are compared.
\end{abstract}

\maketitle


\section{Introduction \label{sec:intro}}
Ultrafast electron diffraction (UED) has demonstrated 
great potential in visualizing the atomic motions in different materials and molecules at the femtosecond timescale~\cite{Filippetto2022review}. UED employs a pump–probe scheme, where a pump laser pulse interacts with the sample and initiates the ultrafast process of interest, followed by a short electron bunch to probe the structural changes over different pump delays. Adjusting the delay between pump laser and probe electron allows time-resolved measurements of the structural evolution of the sample. The temporal resolution of a UED instrument is described by, 
\begin{equation}
    \tau _ { \text {res } } = \sqrt { \tau _ { \text {pump } } ^ { 2 } + \tau _ { \text {probe } } ^ { 2 } + \tau _ { \mathrm { VM } } ^ { 2 } + \tau _ { \text {jitter } } ^ { 2 } }
\end{equation}
where $\tau _ { \text {pump } }$ is the duration of pump laser, $\tau _ { \text {probe } }$ is the duration of electron bunch, $\tau _ { \mathrm { VM } }$ is the velocity mismatch, and $\tau _ { \text {jitter } } $ is the temporal jitter between pump and probe. To improve the temporal resolution, relativistic electrons with energies in the MeV range can be utilized. By employing MeV energy electrons, both bunch lengthening caused by space charge forces and velocity mismatch between pump and probe can be mitigated. Moreover, MeV-UED instruments offer several advantages over their keV counterparts, including providing shorter de Broglie wavelengths and a near flat Ewald's Sphere in the $k$-space, greater penetration depth, as well as reduced background noise due to multiple scattering.

Photocathode radio frequency (RF) guns are often employed in the MeV-UED instruments for the generation of low emittance electron probes with ultrashort duration. The compression of the electron bunch can be achieved by an RF buncher or alternatively by the gun itself. However, fluctuations in the amplitude and phase of RF cavities induce arrival time jitter between pump and probe and contribute to the temporal resolution. To ensure an MeV-UED instrument achieves optimal performance, it is essential to consider all factors including charge, electron bunch length, transverse emittance, and temporal jitter.

While it is common to optimize bunch length and transverse emittance in beamline designs, the optimization of arrival time jitter is not as straightforward. In this work, we present a semi-analytical approach to calculate RF-induced temporal jitter in MeV-UED instruments. Our approach combines a single particle dynamics model~\cite{kim1989} with a differentiable solver, enabling fast and accurate calculation of the exact numerical relation between RF and temporal jitters based on cavity field maps and klystron parameters. The numerical relation can be used to estimate time-of-arrival jitter in an MeV-UED instrument and allows for its online correction from shot-to-shot RF measurements. Compared with previous work~\cite{li2009temporal,alberdi2022novel,cropp2023}, our approach does not require particle-in-cell codes or measured data to determine jitter and applies to beamlines with multiple RF cavities. To showcase our approach, we examine the temporal resolution of the SLAC MeV-UED instrument~\cite{weathersby2015mega} under different beamline configurations in proposed upgrades. The results indicate that RF-based bunch compression with a 1.4 cell gun or a buncher cavity provides shorter bunch duration than a 1.6 cell gun but increased arrival time jitters. Mitigation schemes to improve the temporal resolution are discussed.

\section{RF-induced Timing Jitter\label{sec:source}}

\subsection{Source of Timing Jitter}

\begin{figure*}[t]
\centering
\includegraphics[width=1.5\columnwidth]{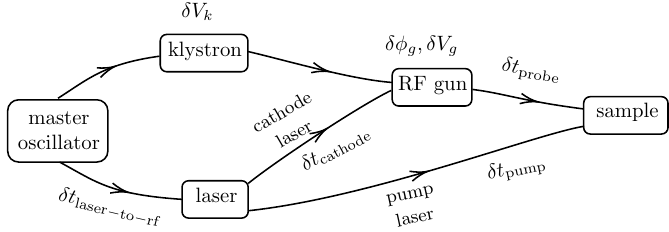}
\caption{Diagram of various sources of timing jitter.}
\label{fig:jitter_diagram}
\end{figure*}

The timing jitter between pump laser and probe electron bunch originates from various sources. Here we consider a beamline with an RF gun as its sole RF element. As depicted in Figure~\ref{fig:jitter_diagram}, the drive signal generated by the master oscillator is split to trigger both klystron amplifier and laser system. On the laser side, the synchronization error between laser and RF produce a timing jitter $\delta t_{\text{laser-to-rf}}$, followed by the transport jitter $\delta t_{\text{cathode}}$ and $\delta t_{\text{pump}}$ in each laser path. The synchronization jitter and transport jitter can be regarded as an $\textit{effective}$ phase jitter in the RF gun $\delta\phi_{l}$ and is given by
\begin{equation}
    \delta \phi_{l} = \omega (\delta t_{\text{laser-to-rf}} + \delta t_{\text{cathode}} - \delta t_{\text{pump}})
\end{equation}
where $\omega$ is the angular frequency of the RF gun. On the RF side, shot-to-shot fluctuation of the klystron voltage $\delta V_{k}$ induces phase and amplitude jitters in the gun. The total phase jitter seen by the electrons is,
\begin{equation}\label{eq:phi_g_tot}
    \delta \phi_{g} =  \delta \hat{\phi}_{g} + \delta\phi_{l}
\end{equation}
here $\delta \hat{\phi}_{g}$ is the phase jitter induced by klystron. The phase jitter $\delta \phi_{g}$ and amplitude jitter $\delta V_{g}$ in the gun translates into arrival time jitter of the probe $\delta t_{\text{probe}}$, which includes the transient time variations in the gun and the drift downstream. Overall, the timing jitter between pump and probe is given by,
\begin{equation}
    \delta t_{\text{sample}} =  \delta t_{\text{probe}} + \delta t_{\text{cathode}} - \delta t_{\text{pump}}
\end{equation}

As we transition to a more quantitative description of jitters, the notation of all jitter quantities will be changed from $\delta$ to $d$ to represent the small deviation from nominal values as the latter is conventionally used in differentiation. We also assume the difference of transport jitters in the pump and probe path is negligible ($\delta t_{\text{cathode}} - \delta t_{\text{pump}} \approx 0$). In practice, this assumption is valid when the laser transport path is short and the temperatures of all optical components are well regulated.

\subsection{Correlated Phase and Amplitude Jitter}
The shot-to-shot fluctuation of amplitude and phase jitter of an RF gun is mainly caused by the fluctuation of klystron voltage and has been well described in the literature~\cite{oh2006,liu2022analysis}. Here we'll give a brief review. The amplitude or voltage $V_{g}$ of an RF gun is related to the klystron voltage $V_{k}$ by
\begin{equation}\label{eq:V_cav}
    V_{g} \left( V _ { k } \right) \propto \sqrt{P_{\text{rf}}} \propto  V _ { k } ^ { 5 / 4 }
\end{equation}
where RF power $P_{\text{rf}}=\nu\mu_{k}V_{k}^{5/2}$, $\nu$ is the efficiency of klystron, $\mu_{k}$ is the perveance. Differentiating both sides of Eq.\ref{eq:V_cav} gives the fractional amplitude jitter of the cavity 
\begin{equation}\label{eq:dV_cav}
    \frac{d V_{g}}{V_{g}} = \frac{5}{4} \frac{d V_{k}}{V_{k}}
\end{equation}

For phase jitters, we assume the phase dependence of the klystron voltage is dominated by transit time of electrons from the input to output cavity. The travel time $t_{k}$ over effective drift length $L_{k}$ in the klystron is,

\begin{equation}\label{eq:tk}
    t_{k} = \frac{L_{k}}{\beta c} =\frac { L _ { k } } { c } \left( 1 - \gamma ^ { - 2 } \right) ^ { - 1 / 2 }
\end{equation}
where $\beta$ is the normalized velocity, $\gamma$ is the Lorentz factor and is related to the klystron voltage $V_{k}$ by,
\begin{equation}\label{eq:gamma}
    \gamma = 1 + \frac{V_{k}}{M}
\end{equation}
here $M=511$~MV is the rest mass voltage of the electron. Inserting Eq.~\ref{eq:gamma} into Eq.~\ref{eq:tk} and converting into phase, we have the klystron output phase $ \phi _ { k } $ and as
\begin{equation}\label{eq:phi_cav}
     \phi _ { k } \left( V _ { k } \right) = 2 \pi f _ { \mathrm { RF } } \frac { L _ { k } } { c } \left[ 1 - \left( 1 + \frac { V _ { k } } { M } \right) ^ { - 2 } \right] ^ { - 1 / 2 }
\end{equation}
here $f_{\text{RF}}$ is the RF frequency. Differentiating Eq.~\ref{eq:phi_cav} gives the phase jitter of the klystron and the RF cavity,
\begin{equation}\label{eq:dphi_cav}
    d  \hat{\phi} _ { g } = d  \phi _ { k } = \mathcal{K} \frac{ d V_{k} } {V_{k}}
\end{equation}
where $\mathcal{K} $ is defined as
\begin{equation}\label{eq:K_cal}
    \mathcal{K} \left( V _ { k } \right) = - 2 \pi f _ { \mathrm { RF } } \frac { L _ { k } } { c }  \frac { V_{k} } { M \left( 1 + \frac { V_{k} } { M } \right) ^ { 3 } \left[ 1 - \frac { 1 } { \left( 1 + \frac { V_{k} } { M } \right) ^ { 2 } } \right] ^ { 3 / 2 } } 
\end{equation}
and quantifies the correlation between phase jitter and klystron voltage jitter.

From Eq.~\ref{eq:dphi_cav}, we can see the amplitude and phase jitter in the RF cavity is related by
\begin{equation}\label{eq:phi_g_hat_corrrelation}
    d  \hat{\phi} _ { g }  = \frac { 4 \mathcal { K } } { 5 } \frac { d V_{g} } { V_{g} }
\end{equation}

In the presence of an active feedback system, the amplitude and phase jitters of the RF gun are reduced from the values in Eq.~\ref{eq:dV_cav} and Eq.~\ref{eq:dphi_cav}. However, the reduction is usually not perfect and the residual jitters still contribute to timing jitters.

\section{Timing Jitter Calculation\label{sec:jitter_calculation}}
In this section we calculate timing jitters downstream of RF cavities from their phase and amplitude jitters. The longitudinal dynamics in a standing-wave RF cavity can be described as~\cite{kim1989},
\begin{eqnarray}\label{eq:dphi_dz}
\frac { \mathrm { d } \phi } { \mathrm { d } z } &=& k ( \frac { \gamma } { \sqrt { \gamma ^ { 2 } - 1 } } - 1)
\end{eqnarray}
\begin{eqnarray} \label{eq:dgamma_dz}
\frac { \mathrm { d } \gamma } { \mathrm { d } z } = 2 \alpha k \mathcal{E}_{z} (z) \sin \left( \phi + k z \right) 
\end{eqnarray}
where $\phi=\omega t - kz+\phi_{i}$, $k$ is the cavity wave number, $\phi_{i}$ is the initial phase of the cavity, $\alpha = \frac { e E _ { 0 } } { 2 m _ { e } c ^ { 2 } k }$ is the normalized field gradient, $\mathcal{E}_{z} (z) $ is the dimensionless normalized field profile peaked at unity. The effect of image charge is ignored as it is verified that the contribution is negligible. The equations above applies to RF guns, linearizer, and buncher cavities with appropriate initial values for $\phi_{i}$ and energy $\gamma_{i}$. For an RF gun, initial phase jitter $d\phi_{i,\mathrm{g}}$ and amplitude jitter $d \alpha_{\mathrm{g}}$ induce a phase and an energy jitter at gun exit and result in a timing jitter at a location downstream,
\begin{equation}\label{eq:dt_tot}
    dt_{\mathrm{tot,g}} (d\phi_{i,\mathrm{g}}, d \alpha_{\mathrm{g}}) = dt_{\mathrm{cav,g}} + dt_{\mathrm{drift,g}}
\end{equation}
where,
\begin{equation}\label{eq:dt_gun}
    dt_{\mathrm{cav,g}} = \frac{1}{\omega} (d \phi_{f,\mathrm{g}} - d \phi_{i,\mathrm{g}})
\end{equation}
\begin{equation}\label{eq:dt_drift}
    dt_{\mathrm{drift,g}} = - \frac{L} {c {\tilde{\beta}_{f,\mathrm{g}}}^{3} {\tilde{\gamma}_{f,\mathrm{g}}}^{3}} d \gamma_{f,\mathrm{g}}
\end{equation}
here $L$ is the drift length after the gun, $\tilde{\beta}_{f,\mathrm{g}}$ and $\tilde{\gamma}_{f,\mathrm{g}}$ are the normalized velocity and Lorentz factor at gun exit in nominal condition, $d \phi_{f,\mathrm{g}}$ and $d \gamma_{f,\mathrm{g}}$ are the phase and energy jitter at cavity exit and are related to the initial jitter $d\phi_{i,\mathrm{g}}$ and $d \alpha_{\mathrm{g}}$ by,


\begin{equation}\label{eq:jacobian_gun}
    \begin{pmatrix}
d\phi_{f,\mathrm{g}}\\
d\gamma_{f,\mathrm{g}}
\end{pmatrix} 
= 
\begin{pmatrix}
\frac{\partial \phi_{f,\mathrm{g}}}{\partial \phi_{i,\mathrm{g}}} &\frac{\partial \phi_{f,\mathrm{g}}}{\partial \alpha_{\mathrm{g}}}\\
\frac{\partial \gamma_{f,\mathrm{g}}}{\partial \phi_{i,\mathrm{g}}} &\frac{\partial \gamma_{f,\mathrm{g}}}{\partial \alpha_{\mathrm{g}}}
\end{pmatrix} 
\begin{pmatrix}
d\phi_{i,\mathrm{g}}\\
d\alpha_{\mathrm{g}}
\end{pmatrix} 
\end{equation}

\begin{figure}[b] 
\centering
\includegraphics[width=1.0\linewidth]{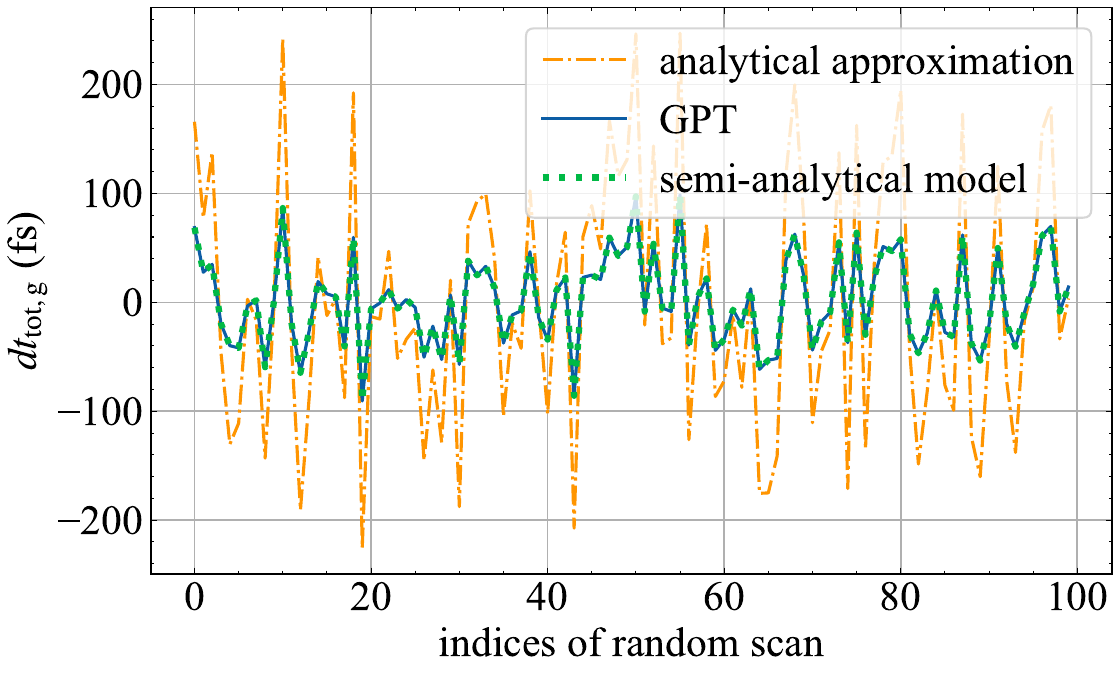}
\caption{Comparison of timing jitters at gun exit obtained by different methods.}
\label{fig:gunmap_and_gpt_verify}
\end{figure}

It is worth pointing out that the $ dt_{\mathrm{tot,g}}$ above does not represent the timing jitter compared with an ideal probe bunch with zero phase, amplitude or laser-to-rf jitter, but rather the timing jitter compared with the pump laser. When there's synchronization error between laser and RF, both the pump and cathode laser will experience laser-to-rf jitter and its contribution needs to be included in $d\phi_{i,\mathrm{g}}$ for the calculation of $ dt_{\mathrm{tot,g}}$.

Kim has derived an analytical solution of $\phi_{f,\mathrm{g}}$  and $\gamma_{f,\mathrm{g}}$ in~\cite{kim1989} for $\mathcal{E}_{z} (z) = \cos(kz)$ which allows the calculation of the Jacobian matrix in Eq.~\ref{eq:jacobian_gun} and the jitters. However, it is an approximate solution and can't determine the timing jitters accurately. Since we are only interested in the jitters around the nominal gun initial phase and gradient, numerical techniques can be used to accurately calculate the partial derivatives around the given point~\cite{baydin2018automatic,berz1999book}. By solving Eq.~\ref{eq:dphi_dz} and Eq.~\ref{eq:dgamma_dz} with a numerical integrator compatible with automatic differentiation~\cite{jax2018github} or differential algebra~\cite{izzo2017differentiable}, one can obtain the final phase and energy as well as their derivatives with respect to the initial phase and gradient. For example, for a 1.5 cell S-band gun with $\mathcal{E}_{z} (z)=\cos{kz}$ operating at $\alpha_{\mathrm{g}}=1.47$ and $\phi_{i,\mathrm{g}}=\pi/6$, the final phase and energy jitter at $z=\frac{3}{4}\lambda$ are related to the initial jitter by,
\begin{equation}\label{eq:dphif_dgammaf_example}
\begin{pmatrix}
d\phi_{f,\mathrm{g}}\\
d\gamma_{f,\mathrm{g}}
\end{pmatrix} 
= 
\begin{pmatrix}
0.65&-0.31\\
1.93&3.78
\end{pmatrix} 
\begin{pmatrix}
d\phi_{i,\mathrm{g}}\\
d\alpha_{\mathrm{g}}
\end{pmatrix} 
\end{equation}
Inserting it back into Eq.~\ref{eq:dt_tot} gives a linear relation between timing jitter downstream and initial RF jitters. It has been shown in~\cite{cropp2023} that a data-driven model based on linear regression can accurately predict shot-to-shot fluctuation in time of arrival, so our model is validated by the observation. Derivatives of higher orders can be included for larger fluctuation of phase and amplitude, which effectively yields a Taylor map between initial RF jitter and final timing jitter. In addition, the semi-analytical model allows extraction of derivatives for cavities with realistic field maps by incorporating a numerical interpolator; see~\cite{codeexample} for a {\sc python} implementation. Jitter calculations for a 1.6 cell and 1.4 cell gun will be presented in Sec.~\ref{sec:beamline_study}.

We compare the timing jitters at exit ($z$=0.1~m) of the 1.5 cell S-band RF gun obtained by different methods. For benchmark, we generate a 10~fC electron beam distribution with 100~$\mu$m radius and 60~fs FWHM duration and track its time of arrival in the same RF gun field with {\sc gpt}~\cite{gpt}. 100 simulations with different random realizations of the RF amplitude and phase were performed and the fractional amplitude and phase jitter values were randomly generated with a normal distribution with respective rms values of 500 ppm and 0.1 degree. As a benchmark, the magnitude of jitter values are larger than typically achieved values. As depicted in Figure~\ref{fig:gunmap_and_gpt_verify}, the analytical approximate solution overestimates the magnitude of timing jitters while the jitters calculated with semi-analytical model agree well with the {\sc gpt} results (maximum deviation within 0.1~fs).

\begin{figure*}[!t] 
\centering
\includegraphics[width=1.9\columnwidth]{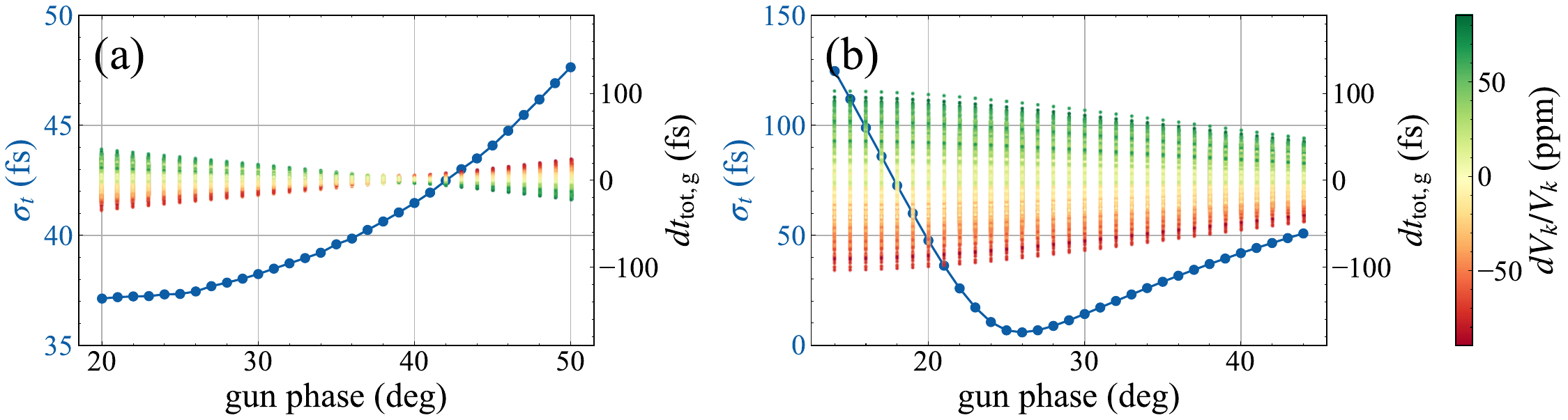}
\caption{Bunch length and arrival time jitter at sample plane downstream of the 1.6 cell gun (a) and 1.4 cell gun (b) for different initial phases. The blue traces denote the evolution of rms bunch length for different gun phases. The colored scatter plots denote the timing jitters corresponding to different klystron jitters.}
\label{fig:1.6-1.4-compare}
\end{figure*}

For calculations involving cascaded cavities, i.e., a buncher at the downstream of an RF gun, the dependence on upstream energy jitter needs to be included, 
\begin{equation}\label{eq:jacobian_buncher}
    \begin{pmatrix}
d\phi_{f,\mathrm{b}}\\
d\gamma_{f,\mathrm{b}}
\end{pmatrix} 
= 
\begin{pmatrix}
\frac{\partial \phi_{f,\mathrm{b}}}{\partial \phi_{i,\mathrm{b}}} &\frac{\partial \phi_{f,\mathrm{b}}}{\partial \alpha_{\mathrm{b}}} &\frac{\partial \phi_{f,\mathrm{b}}}{\partial \gamma_{f,\mathrm{g}}} \\
\frac{\partial \gamma_{f,\mathrm{b}}}{\partial \phi_{i,\mathrm{b}}} &\frac{\partial \gamma_{f,\mathrm{b}}}{\partial \alpha_{\mathrm{b}}} &\frac{\partial \gamma_{f,\mathrm{b}}}{\partial \gamma_{f,\mathrm{g}}}
\end{pmatrix} 
\begin{pmatrix}
d \phi_{i,\mathrm{b}}\\
d\alpha_{\mathrm{b}} \\
d\gamma_{f,\mathrm{g}}
\end{pmatrix} 
\end{equation}
here $d \phi_{i,\mathrm{b}}$ is the total buncher phase jitter analogous to Eq.~\ref{eq:phi_g_tot} and includes the phase jitter $d\hat{\phi}_{i,\mathrm{b}} $ from the buncher cavity, synchronization jitter between laser and RF, and arrival time jitter at buncher entrance $dt_{\text{tot,g}}$. The timing jitter downstream of the buncher can be calculated similarly as Eq.~\ref{eq:dt_tot}.

Given the layout, cavity field maps, and operating parameters of a beamline, the numerical relation between RF and temporal jitter can be obtained from Eq.~\ref{eq:dt_tot}, with the Jacobian matrix in Eq.~\ref{eq:jacobian_gun} and Eq.~\ref{eq:jacobian_buncher} calculated from a differentiable solver. The magnitude of temporal jitter can then be estimated based on performance of the klystron and laser-to-RF synchronization. In the following section, we'll use the approach to study the impact of RF jitters in the SLAC MeV-UED instrument.

The single particle model described above can also be used to describe evolution of bunch duration due to RF effect~\cite{li2009temporal}. Considering two particles emitted at time $-\frac{\tau_{i}}{2}$ and $\frac{\tau_{i}}{2}$, the final bunch duration is given by,
\begin{equation}
    \tau_{f} = \tau_{i} + dt_{\mathrm{tot,g}}(\frac{\tau_{i}}{2\omega}, 0) - dt_{\mathrm{tot,g}}(-\frac{\tau_{i}}{2\omega}, 0)
\end{equation}
For strongly compressed beams, longitudinal space charge plays an important role near the waist position and the single particle model fails to predict the correct bunch duration near the waist. A more thorough treatment is needed to include this scenario~\cite{denham2023} and is beyond the scope of this work.

\section{Temporal Resolution of the SLAC MeV-UED Instrument \label{sec:beamline_study}}

\subsection{1.6 Cell and 1.4 Cell Gun}\label{subsec:gun_study}
The SLAC MeV-UED instrument has been developing robust methods to generate ultrashort electron bunches with low arrival time jitters~\cite{weathersby2015mega}. Currently, an LCLS-type 1.6 cell S-band photocathode rf gun serves as the electron source~\cite{Akre2008}. A 1.4 cell gun is being considered as the next generation electron source as it provides higher extraction field and bunch compression~\cite{pirez2017s,song2022development}. 

\begin{figure}[t] 
\centering
\includegraphics[width=1.0\linewidth]{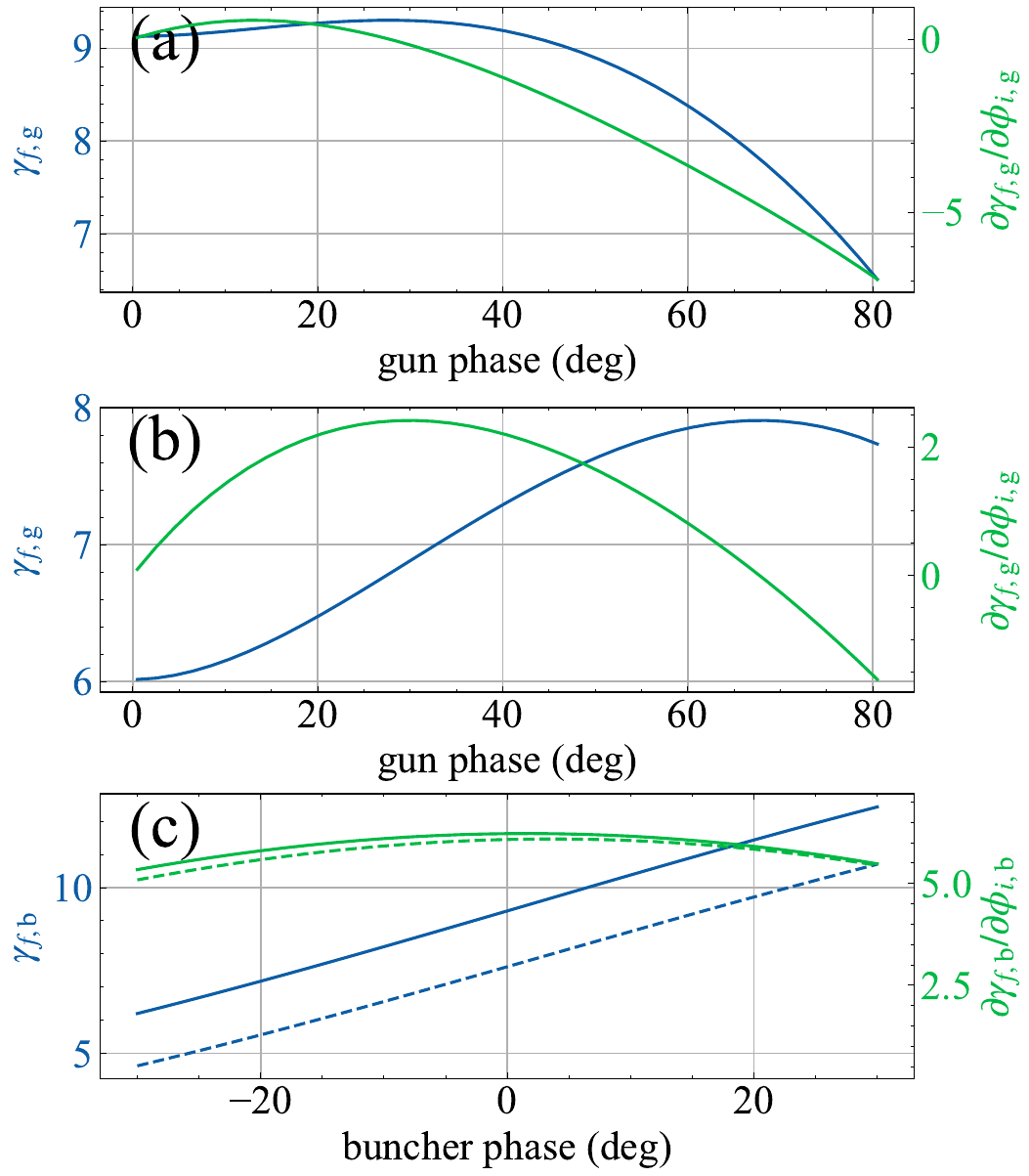}
\caption{Exit energy, and its slope w.r.t phase (in radian) for different initial phases of 1.6 cell gun (a), 1.4 cell gun ((b), 1.6 cell gun and buncher (c). The solid and dashed trace correspond to gun initial phases at 30 and 70 degrees.}
\label{fig:phase_scan_gamma}
\end{figure}

Here we compare the bunch length and timing jitter of the electron bunches produced from the 1.6 cell and 1.4 cell gun for different initial phases at $z=1.55$m (sample location). Both guns are assumed to operate at nominal gradients of 90 MV/m with $\alpha_{\mathrm{g}}=1.47$ . The timing jitters are calculated using our semi-analytical method, and the bunch length is modeled with particle tracking code {\sc gpt}. For the jitter sources, we assume the fractional klystron voltage jitter to be 20~ppm (rms), and laser-to-rf synchronization jitter to be 10~fs (rms)~\cite{ma2019slac}. The correlation coefficient between phase and klystron jitter $\mathcal{K}$ is -14.6 based on specification of the ScandiNova klystron system currently in use at the SLAC facility. The effects of RF feedback systems are not included in the jitter calculation and therefore the jitter values should be slightly overestimated.

\begin{figure*}[t] 
\centering
\includegraphics[width=1.8\columnwidth]{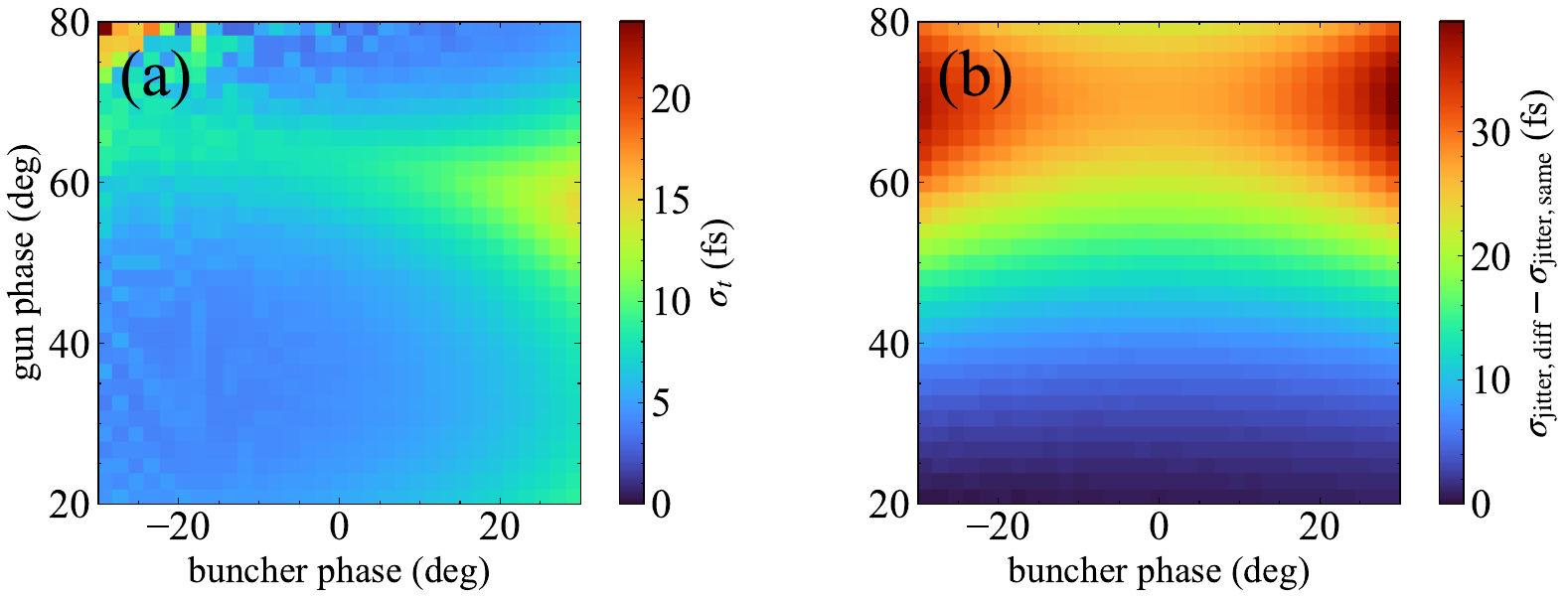}
\caption{(a) Minimum bunch lengths achieved at different gun and buncher phases. (b) Difference of timing jitters when the two cavities are powered by different and same klystrons.}
\label{fig:buncher_jitter_scan}
\end{figure*}

The results of the timing jitter and bunch length simulation are shown in Figure~\ref{fig:1.6-1.4-compare}. The 1.6 cell gun doesn't provide significant bunch compression and the bunch length continues to grow when increasing the launch phase. This can be understood by looking at the energy dependence of the launch phase as shown in Figure~\ref{fig:phase_scan_gamma}(a).  The tail particles (higher phase) don't gain much more energy than the head particles and therefore can't catch up in velocity bunching. The slope of the energy-phase curve flips at higher emission phases and the bunch will be lengthened. During the transition, there exists a nominal launch phase that minimizes the temporal jitter. Using the approach from Sec.~\ref{sec:jitter_calculation}, at the launch phase of 38 degree, the relation between timing jitter and amplitude jitter and phase jitter can be expressed as,
\begin{equation}
    dt_{\mathrm{g,tot}} (\mathrm{ps}) = -53.69 d\alpha_{\mathrm{g}} - 6.574 d\phi_{i,\mathrm{g}}
\end{equation}
Since the phase jitter induced by klystron $d\hat{\phi}_{i,\mathrm{g}}$ is negatively correlated to the amplitude jitter by Eq.~\ref{eq:phi_g_hat_corrrelation}, a large part of the RF-induced timing jitter is cancelled. Similar phenomenon exists in linear accelerators where a certain post-crest acceleration angle can minimize klystron-induced energy variations~\cite{Mansten2024}.

The 1.4 cell gun provides a much stronger bunch compression than the 1.6 cell gun as the energy gain increases with the initial phase with a larger slope. For the desired sample location, the minimum bunch duration achieved is 5 fs when operating at an initial phase of 26 degree. However, the stronger bunch compression is accompanied by larger timing jitters. As shown in Figure~\ref{fig:phase_scan_gamma}(b), when operating at linear part of the energy-phase curve, the exit energy is sensitive to initial phases jitters. The energy jitters are then converted into timing jitters in the drift after the gun. 

Assuming negligible velocity mismatch between pump and probe, and including pump laser duration (25 fs rms), electron bunch length and timing jitter, the 1.6 cell gun achieves an overall temporal resolution of 46 fs (rms) when operating at 25 degree. The optimal phase for temporal resolution deviates from the phase of minimum jitter (around 40 degree) since the temporal resolution is dominated by the electron bunch length. For the 1.4 cell gun, the optimal temporal resolution is 33 fs when operating at 27 degree and the dominant contribution is the timing jitter and pump laser duration.

\subsection{Gun and Buncher}

RF buncher provides another path to bunch compression and improvement of temporal resolution. As discussed in the previous section, a stronger bunch compression can result in larger timing jitter. In this scenario, since there are two RF cavities in the beamline, this raises the question of whether using the same klystron for both the gun and buncher could reduce overall timing jitter and provide benefits~\cite{franssen2017improving}.

We consider a beamline with a 1.6 cell gun followed by a 5-cell S band buncher operating at 25 MV/m at $z=0.75$m from the cathode. We scan the initial phases for gun and buncher and perform particle-in-cell simulation to find the minimum bunch lengths for each initial phases and the corresponding waist locations. The timing jitters at waist locations are calculated using the klystron and synchronization parameters in Sec.~\ref{subsec:gun_study}, assuming the gun and buncher are powered by the same klystron---the fractional amplitude and phase jitters for the two cavities are identical. The calculation is then repeated for the scenario where the RF phase and amplitude jitters for gun and buncher are independent. The results of the minimum bunch lengths and the difference in timing jitters for the two cases are shown in Figure~\ref{fig:buncher_jitter_scan}. For the same-klystron configuration, the rms timing jitters at waist do not show clear dependence on the initial phases of gun and buncher and are around 18~fs for the phase range considered. The dependence is more prominent for the different-klystron configuration and timing jitter is larger when operating at higher gun phases. The advantage of the same-klystron configuration only becomes apparent at higher gun phases.

The mechanism of jitter reduction in a gun-buncher beamline can be understood in two aspects. The timing jitter at buncher entrance $dt_{\mathrm{tot,g}}$ is generally suppressed after the buncher due to the process of bunch compression. This reduction happens regardless of whether the gun and buncher are powered by the same klystron or not. The energy jitter $d\gamma_{f,\mathrm{g}}$ can be compensated in the same-klystron configuration at higher gun phases as the dependence of energy with w.r.t phase, $\frac{\partial \gamma_{f,\mathrm{g}}}{\partial \phi_{i,\mathrm{g}}}$ and $\frac{\partial \gamma_{f,\mathrm{b}}}{\partial \phi_{i,\mathrm{b}}}$, have opposite signs and are close in absolute values (see Figure~\ref{fig:phase_scan_gamma} (c)). The energy jitter at buncher exit, and the time-of-flight jitter downstream of the buncher, is then reduced in this operating regime.

\begin{table}[h]
    \centering
    \begin{tabular}{c|c|c|c}
         & $\sigma_{t}$ (fs) & $\sigma_{\text{jitter}}$ (fs) & $\sigma_{\text{res}}$ (fs) \\ \hline
       1.6 cell gun  & 37 & 6 & 46\\ \hline
       1.4 cell gun  & 7 & 20 & 33 \\ \hline
       1.6 cell gun + buncher  &  6&  18& 31 \\ \hline
       1.4 cell gun + buncher  &  4&  18 & 31\\ \hline
    \end{tabular}
    \caption{RMS bunch length, arrival time jitter, and temporal resolution of different beamline configurations. The temporal resolution is calculated including the contribution of pump laser duration $\sigma_{\text{pump}}$=25 fs and assumes klystron voltage jitter of 20ppm (rms) and laser-to-rf synchronization jitter of 10fs (rms).}
    \label{tab:temporal_res}
\end{table}

The optimal temporal resolution achieved by the 1.6 cell gun and the buncher is around 31 fs for the two configurations (gun and buncher powered by same or different klystrons). We also considered a beamline with the 1.4 cell gun and the buncher and found the level of arrival time jitter is similar. For brevity, the corresponding graphs are not included. The rms bunch duration, arrvial time jitter and temporal resolution of a 1.4 cell gun and buncher beamline are summarized in Table~\ref{tab:temporal_res} along with other beamline configurations. Overall, the main contribution to temporal resolution in a gun-and-buncher beamline is the arrival time jitter and finite pump duration. 

\subsection{Jitter Correction}

\begin{figure}[h] 
\centering
\includegraphics[width=0.9\linewidth]{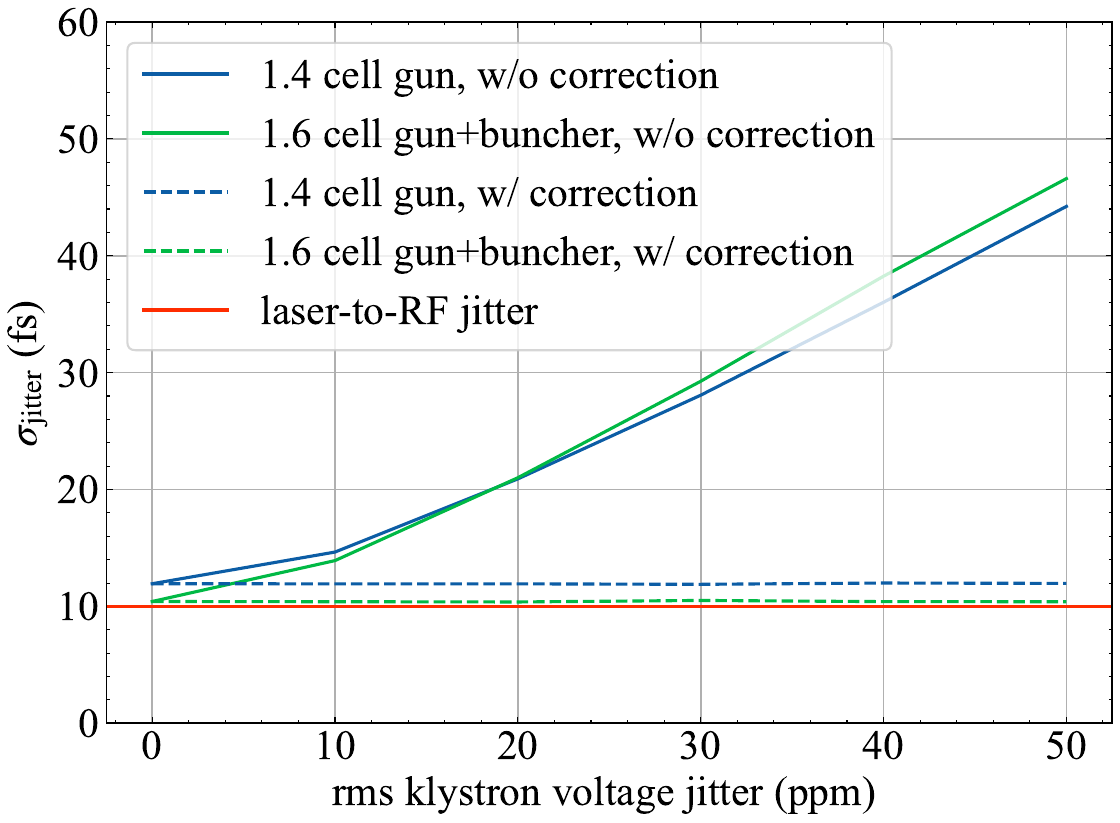}
\caption{Comparison of timing jitters with and without jitter correction for 1.4 cell gun and buncher.}
\label{fig:jitter_correction}
\end{figure}

When the temporal resolution is dominated by the timing jitters, the approach outlined in Section~\ref{sec:jitter_calculation} serves as a virtual timing tool to predict the RF-induced timing jitter. Since RF amplitude and phase jitters are collected every shot, the measured RF jitter values allow direct calculation of timing jitters with our approach. The predicted time jitter values can be used to sort the corresponding diffraction data and partially correct the arrival time jitters---the effect of laser-to-rf synchronization error remains and cannot be corrected.

We evaluate the correction approach by considering the timing jitters of the 1.4 cell gun beamline and the 1.6 cell gun and buncher beamline from previous sections. The 1.4 cell gun is assumed to operate at 26 deg, while the 1.6 cell gun and buncher operate at 30 deg phase and zero crossing respectively and are powered by different klystrons. We vary the level of klystron voltage jitter and calculate the rms timing jitter with and without correction. The rms value of laser-to-rf locking error is assumed to be 10~fs. The results of the calculation are shown in Figure~\ref{fig:jitter_correction}. The correction scheme reduces the jitters across different klystron jitter levels, with the residual jitter limited by laser-to-rf synchronization performance. To minimize the residual jitter, a THz streaking structure can be used to directly measure the time jitters relative to pump laser~\cite{zhao2018,li2019,othman2024improved}. 

\section{Conclusion}
In summary, we have presented a semi-analytical method for calculating RF-induced timing jitters from cavity and klystron parameters. The method is based on a single particle dynamics model and directly connects the RF and temporal jitters via simple numerical relations. The semi-analytical model yields good agreement with results from particle-in-cell simulation and allows fast evaluation of temporal resolution in MeV-UED beamlines.

To demonstrate our method, we assessed the temporal resolution of the SLAC MeV-UED instrument under different configurations. Specifically, we show that operating the 1.6 cell gun at a nominal launch phase can minimize the timing jitter due to the correlation between klystron phase and amplitude jitter. The performance of a proposed 1.4 cell gun is examined and we found that the overall temporal resolution achieved is dominated by the temporal jitter. Additionally, we considered a beamline with a 1.6 cell gun and an RF buncher to improve temporal resolution. The timing jitters are compared for the two scenarios where the gun and buncher are powered by same or different klystrons, and we showed that the same-klystron configuration can achieve reduced jitter especially at higher gun launch phases.

Improvement of temporal resolution in an MeV-UED instruments requires minimizing both electron bunch duration and timing jitters. Our proposed method enables rapid calculation of timing jitters and can be integrated with particle tracking code to optimize the overall performance of UED beamlines. For example, optimization of RF phase and amplitudes of multiple cavities could lead to simultaneous improvement of bunch duration and timing jitters~\cite{franssen2017improving,alberdi2022novel}. The operating phase of the RF gun can also be chosen to minimize the timing jitters at gun exit, with downstream jitters compensated by a magnetic compressor~\cite{kim_towards_2020,qi2020}. When the timing jitter is dominated by the contribution from klystron voltage variation, our approach also allows shot-to-shot calculation of time-of-arrival for online jitter correction.

\section{Acknowledgements}
We thank Lili Ma (SLAC) for useful discussion on laser-to-rf synchronization. This work was supported by the U.S. Department of Energy Office of Science, Office of Basic Energy Sciences under Contract No. DE-AC02-76SF00515. The SLAC MeV-UED program is supported by the U.S. Department of Energy Office of Science, Office of Basic Energy Science under FWP 10075, FWP 100713, and FWP 100940.

\bibliographystyle{ieeetr}
\bibliography{sample}

\begin{thebibliography}{10}

\bibitem{Filippetto2022review}
D.~Filippetto, P.~Musumeci, R.~K. Li, B.~J. Siwick, M.~R. Otto, M.~Centurion,
  and J.~P.~F. Nunes, ``Ultrafast electron diffraction: Visualizing dynamic
  states of matter,'' {\em Rev. Mod. Phys.}, vol.~94, p.~045004, Dec 2022.

\bibitem{kim1989}
K.-J. Kim, ``Rf and space-charge effects in laser-driven rf electron guns,''
  {\em Nucl. Instrum. Methods Phys. Res., Sect. A}, vol.~275, no.~2,
  pp.~201--218, 1989.

\bibitem{li2009temporal}
R.-K. Li and C.-X. Tang, ``Temporal resolution of {MeV} ultrafast electron
  diffraction based on a photocathode rf gun,'' {\em Nucl. Instrum. Methods
  Phys. Res., Sect. A}, vol.~605, no.~3, pp.~243--248, 2009.

\bibitem{alberdi2022novel}
B.~Alberdi~Esuain, J.-G. Hwang, A.~Neumann, and T.~Kamps, ``Novel approach to
  push the limit of temporal resolution in ultrafast electron diffraction
  accelerators,'' {\em Scientific Reports}, vol.~12, no.~1, p.~13365, 2022.

\bibitem{cropp2023}
F.~Cropp, L.~Moos, A.~Scheinker, A.~Gilardi, D.~Wang, S.~Paiagua, C.~Serrano,
  P.~Musumeci, and D.~Filippetto, ``Virtual-diagnostic-based time stamping for
  ultrafast electron diffraction,'' {\em Phys. Rev. Accel. Beams}, vol.~26,
  p.~052801, May 2023.

\bibitem{weathersby2015mega}
S.~Weathersby, G.~Brown, M.~Centurion, T.~Chase, R.~Coffee, J.~Corbett,
  J.~Eichner, J.~Frisch, A.~Fry, M.~G{\"u}hr, {\em et~al.},
  ``Mega-electron-volt ultrafast electron diffraction at slac national
  accelerator laboratory,'' {\em Review of Scientific Instruments}, vol.~86,
  no.~7, 2015.

\bibitem{oh2006}
J.-S. Oh, T.~Hara, T.~Inagaki, T.~Shintake, and K.~Shirasawa, ``{Stable RF
  Phase Insensitive to the Modulator Voltage Fluctuation of the C-band Main
  Linac for SCSS XFEL},'' in {\em Proc. FEL'06}, pp.~684--687, JACoW
  Publishing, Geneva, Switzerland, 2006.

\bibitem{liu2022analysis}
Y.~Liu, H.~Matsumoto, L.~Li, and M.~Gu, ``Analysis the influence of
  pulse-to-pulse stability of modulator on high-power microwave output of
  pulsed klystron,'' {\em SN Applied Sciences}, vol.~4, no.~1, p.~4, 2022.

\bibitem{baydin2018automatic}
A.~G. Baydin, B.~A. Pearlmutter, A.~A. Radul, and J.~M. Siskind, ``Automatic
  differentiation in machine learning: a survey,'' {\em Journal of machine
  learning research}, vol.~18, no.~153, pp.~1--43, 2018.

\bibitem{berz1999book}
M.~Berz, {\em Modern map methods in particle beam physics}, vol.~108.
\newblock Academic Press, 1999.

\bibitem{jax2018github}
J.~Bradbury, R.~Frostig, P.~Hawkins, M.~J. Johnson, C.~Leary, D.~Maclaurin,
  G.~Necula, A.~Paszke, J.~Vander{P}las, S.~Wanderman-{M}ilne, and Q.~Zhang,
  ``{JAX}: composable transformations of {P}ython+{N}um{P}y programs,'' 2018.

\bibitem{izzo2017differentiable}
D.~Izzo, F.~Biscani, and A.~Mereta, ``Differentiable genetic programming,'' in
  {\em Genetic Programming: 20th European Conference, EuroGP 2017, Amsterdam,
  The Netherlands, April 19-21, 2017, Proceedings 20}, pp.~35--51, Springer,
  2017.

\bibitem{codeexample}
``https://github.com/alfven17/ued-timing-jitter.''
  \url{https://github.com/Alfven17/ued-timing-jitter}.
\newblock [Code example for calculating Jacobian matrix].

\bibitem{gpt}
``{General Particle Tracer}.'' \url{https://www.pulsar.nl/gpt/index.html}.

\bibitem{denham2023}
P.~Denham and P.~Musumeci, ``Analytical scaling laws for radiofrequency-based
  pulse compression in ultrafast electron diffraction beamlines,'' {\em
  Instruments}, vol.~7, no.~4, 2023.

\bibitem{Akre2008}
R.~Akre, D.~Dowell, P.~Emma, J.~Frisch, S.~Gilevich, G.~Hays, P.~Hering,
  R.~Iverson, C.~Limborg-Deprey, H.~Loos, A.~Miahnahri, J.~Schmerge, J.~Turner,
  J.~Welch, W.~White, and J.~Wu, ``Commissioning the linac coherent light
  source injector,'' {\em Phys. Rev. ST Accel. Beams}, vol.~11, p.~030703, Mar
  2008.

\bibitem{pirez2017s}
E.~Pirez, P.~Musumeci, J.~Maxson, and D.~Alesini, ``S-band 1.4 cell
  photoinjector design for high brightness beam generation,'' {\em Nucl.
  Instrum. Methods Phys. Res., Sect. A}, vol.~865, pp.~109--113, 2017.

\bibitem{song2022development}
Y.~Song, J.~Yang, J.~Wang, J.~Urakawa, T.~Takatomi, and K.~Fan, ``Development
  of a 1.4-cell rf photocathode gun for single-shot mev ultrafast electron
  diffraction devices with femtosecond resolution,'' {\em Nucl. Instrum.
  Methods Phys. Res., Sect. A}, vol.~1031, p.~166602, 2022.

\bibitem{ma2019slac}
L.~Ma, X.~Shen, K.~Kim, D.~Brown, M.~D'Ewart, B.~Hong, J.~Olsen, S.~Smith,
  D.~Van~Winkle, E.~Williams, {\em et~al.}, ``Slac ued llrf system upgrade,''
  {\em arXiv preprint arXiv:1910.02296}, 2019.

\bibitem{Mansten2024}
E.~Mansten, R.~Sv\"ard, S.~Thorin, M.~Eriksson, and P.~F. Tavares,
  ``Cancellation of klystron-induced energy and arrival-time variations in
  linear accelerators with arc-type bunch compressors,'' {\em Phys. Rev. Accel.
  Beams}, vol.~27, p.~040401, Apr 2024.

\bibitem{franssen2017improving}
J.~Franssen and O.~Luiten, ``Improving temporal resolution of ultrafast
  electron diffraction by eliminating arrival time jitter induced by
  radiofrequency bunch compression cavities,'' {\em Structural Dynamics},
  vol.~4, no.~4, 2017.

\bibitem{zhao2018}
L.~Zhao, Z.~Wang, C.~Lu, R.~Wang, C.~Hu, P.~Wang, J.~Qi, T.~Jiang, S.~Liu,
  Z.~Ma, F.~Qi, P.~Zhu, Y.~Cheng, Z.~Shi, Y.~Shi, W.~Song, X.~Zhu, J.~Shi,
  Y.~Wang, L.~Yan, L.~Zhu, D.~Xiang, and J.~Zhang, ``Terahertz streaking of
  few-femtosecond relativistic electron beams,'' {\em Phys. Rev. X}, vol.~8,
  p.~021061, Jun 2018.

\bibitem{li2019}
R.~K. Li, M.~C. Hoffmann, E.~A. Nanni, S.~H. Glenzer, M.~E. Kozina, A.~M.
  Lindenberg, B.~K. Ofori-Okai, A.~H. Reid, X.~Shen, S.~P. Weathersby, J.~Yang,
  M.~Zajac, and X.~J. Wang, ``Terahertz-based subfemtosecond metrology of
  relativistic electron beams,'' {\em Phys. Rev. Accel. Beams}, vol.~22,
  p.~012803, Jan 2019.

\bibitem{othman2024improved}
M.~A. Othman, A.~E. Gabriel, E.~C. Snively, M.~E. Kozina, X.~Shen, F.~Ji,
  S.~Lewis, S.~Weathersby, P.~Vasireddy, D.~Luo, {\em et~al.}, ``Improved
  temporal resolution in ultrafast electron diffraction measurements through
  thz compression and time-stamping,'' {\em Structural Dynamics}, vol.~11,
  no.~2, 2024.

\bibitem{kim_towards_2020}
H.~W. Kim, N.~A. Vinokurov, I.~H. Baek, K.~Y. Oang, M.~H. Kim, Y.~C. Kim, K.-H.
  Jang, K.~Lee, S.~H. Park, S.~Park, J.~Shin, J.~Kim, F.~Rotermund, S.~Cho,
  T.~Feurer, and Y.~U. Jeong, ``Towards jitter-free ultrafast electron
  diffraction technology,'' {\em Nature Photonics}, vol.~14, pp.~245--249, Apr.
  2020.

\bibitem{qi2020}
F.~Qi, Z.~Ma, L.~Zhao, Y.~Cheng, W.~Jiang, C.~Lu, T.~Jiang, D.~Qian, Z.~Wang,
  W.~Zhang, P.~Zhu, X.~Zou, W.~Wan, D.~Xiang, and J.~Zhang, ``Breaking 50
  femtosecond resolution barrier in mev ultrafast electron diffraction with a
  double bend achromat compressor,'' {\em Phys. Rev. Lett.}, vol.~124,
  p.~134803, Mar 2020.

\end{thebibliography}
\end{document}